\documentclass[conference,a4paper]{IEEEtran}

\usepackage[T1]{fontenc}
\usepackage[utf8]{inputenc}

\usepackage{cite}
\usepackage[cmex10]{amsmath}
\usepackage{amssymb,amsfonts}
\usepackage{array}
\usepackage{dblfloatfix}
\usepackage{float}
\usepackage{subfigure}
\usepackage[ruled,vlined]{algorithm2e}
\usepackage{graphicx}
\graphicspath{{graphics/}{moregraphics/}}
\DeclareGraphicsExtensions{.pdf,.PDF,.jpeg,.JPEG,.jpg,.JPEG,.png,.PNG}

\usepackage{booktabs}

\usepackage{siunitx}

\usepackage{textcomp}
\usepackage{xcolor}
\usepackage[spaces,hyphens]{url}
\usepackage{diagbox}
\usepackage{makecell}

\def\BibTeX{{\rm B\kern-.05em{\sc i\kern-.025em b}\kern-.08em
    T\kern-.1667em\lower.7ex\hbox{E}\kern-.125emX}}

\begin{document}

\title{Impact of Intervals on the Emotional Effect in Western Music}

\author{\IEEEauthorblockN{Cengiz Kaygusuz}
\IEEEauthorblockA{Florida International University\\
ckayg001@fiu.edu}
\and
\IEEEauthorblockN{Julian Zuluaga}
\IEEEauthorblockA{Florida International University \\
jzulu013@fiu.edu}
}

\maketitle
\begin{abstract}
Every art form ultimately aims to invoke an emotional response over the audience, and music is no different. While the precise perception of music is a highly subjective topic, there is an agreement in the "feeling" of a piece of music in broad terms. Based on this observation, in this study, we aimed to determine the emotional feeling associated with short passages of music; specifically by analyzing the melodic aspects. We have used the dataset put together by Eerola et. al. which is comprised of labeled short passages of film music. Our initial survey of the dataset indicated that other than "happy" and "sad" labels do not possess a melodic structure. We transcribed the main melody of the happy and sad tracks, and used the intervals between the notes to classify them. Our experiments has shown that treating a melody as a bag-of-intervals do not possess any predictive power whatsoever, whereas counting intervals with respect to the key of the melody yielded a classifier with 85\% accuracy.
\end{abstract}

\section{Introduction}\label{sec:intro1}

The emotional touch of the music is undoubtedly experienced by many, and it is not a coincidence since every form of art is usually intended to evoke a certain emotion on the listener. To be more specific, recent research indicates that music in particular, serves to regulate mood and increase one's self-awareness, clearly establishing the relationship between music and human emotions\cite{gabrielsson2001influence}.

Music is not an atomic construct on its own accord however, and considering its multidimensional essence makes way to a natural question: what is the relationship between the aspects of music and its emotional content? There is a tremendous amount of deliberation put forward for the answer~\cite{scherer2001emotional}, and there seems to be a strong agreement on the emotional effects of various musical phenomena, as well as the impact of the state of the listener.

One aspect of music, namely the \textit{melody}, constitutes a quantifiable measure. It is also the most significant aspect, as listeners mainly relate to musical works by their melodies. In western music, the distance between two notes is measured by "semitones", which can be thought of as the building blocks of a melody. The group of notes in a melody is characterized by the relative distances between them. Western music is comprised of 12 intervals, each one of them having a unique feeling associated to it which can be divided in two, as "happy, energetic, positive" or as "sad, sentimental, negative". We hypothesize that the positive intervals occur more on music that induce happiness, and similarly, negative intervals occur more on music that induce sadness.

\section{Background}

\begin{table}[]
    \centering
    \caption{Musical aspects and their feelings \cite{gabrielsson2001influence}.}
    \label{tab:aspects-feelings}

    \begin{tabular}{| c | c | c |}
        \hline
        \textbf{Aspect} & \textbf{Description} & \textbf{Feeling}  \\ \hline \hline 
        Tempo & Pace of the music & \makecell{Fast: Happiness, excitement, anger.  \\ Slow: Sadness, serenity}.  \\ \hline
        Mode & Type of the scale & \makecell{Major: happiness, joy. \\ Minor: sadness, darkness.} \\ \hline
        Rhythm & Beat of a song & \makecell {Smooth rhythm: happiness, peace. \\ Irregular rhythm: amusement, uneasiness. \\ Varied rhythm: joy.} \\ \hline
    \end{tabular}
\end{table}

Aside from the data mining dimension, this study requires some knowledge of human psychology and music theory, which we aim to present in this section.

\subsection{Emotions, Mood, and Music}
\begin{figure*}[t]
    \centering
    \subfigure[]{
        \includegraphics[scale=0.30]{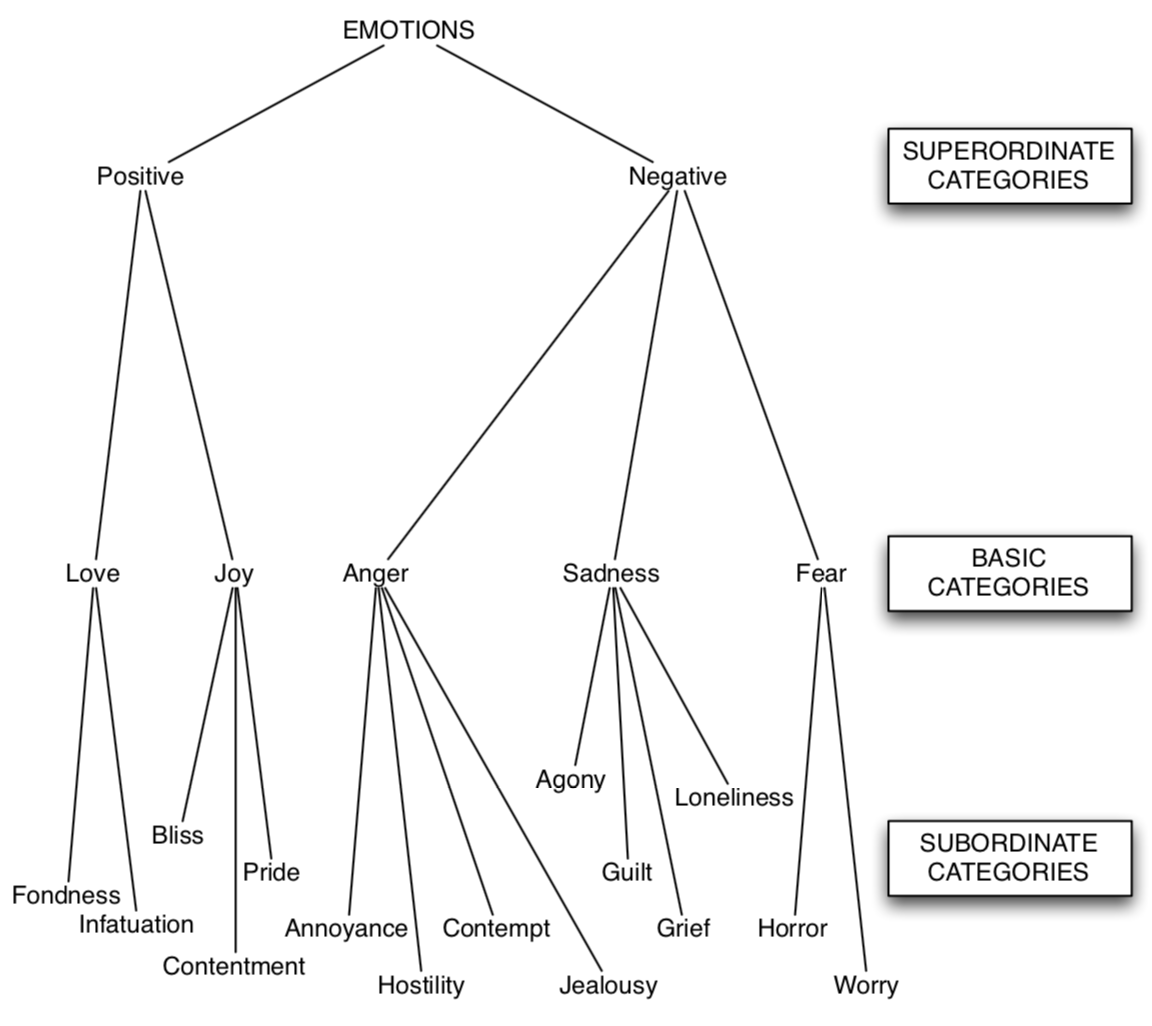}
        \label{fig:hierarchical-taxonomy}
    }
    \qquad 
    \subfigure[]{
        \includegraphics[scale=0.43]{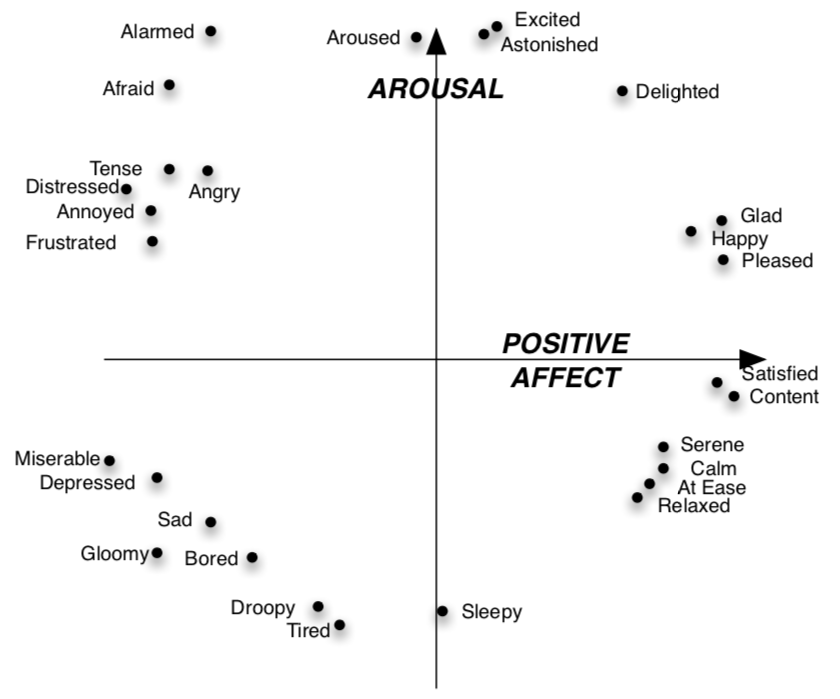}
        \label{fig:dimensional-taxonomy}
    }
    \caption{A hierarchical (a), and a dimensional (b) examples of emotional model \cite{russell1980circumplex, shaver1987emotion}.}
    \label{fig:bucket-size}
\end{figure*}

The concepts of \textit{emotion} and \textit{mood} are closely related to each other as both can be described as \textit{feelings}. However, there are subtle differences between the two, which are not obvious at first sight. The dictionary definition of emotion is "a particular feeling that characterizes the state of mind", whereas mood is defined as "a prevailing emotional tone or general attitude". As their definitions indicate, we can consider emotions as concrete feelings, such as joy, anger, and fear, whereas mood is the \textit{influencer} of emotions, such as the state of being \textit{irritable}. It is with no doubt that music influences both emotions and the mood, however we are often aware of our emotions and seldom acknowledge our mood, hence emotions are much more reliable to reason about.

A basic taxonomy of emotions is given in Figure~\ref{fig:hierarchical-taxonomy}, as given by Schaver\cite{shaver1987emotion}. In this simple hierarchical model, emotions are classified primarily as positive and negative, and get more and more specific as we traverse the tree. 

Other prevalent taxonomies use a dimensional model to classify emotions, basically breaking each emotion into their "essence", which correspond to a dimension in the model. An example two-dimensional model is given in Figure~\ref{fig:dimensional-taxonomy} taken from Russell et. al. \cite{russell1980circumplex}. Other dimensional models vary this basic idea by using different or more axes~\cite{Li:2011:MDM:2016688}. The dataset we used are labeled in both 6 basic emotions, as well as in a 3D model of emotions with the axes named as: valence, activity, and intensity. 

The representability of all kinds of emotion by music is an open question~\cite{Li:2011:MDM:2016688}. However, a few musical aspects are well known to induce specific feelings, given in Table~\ref{tab:aspects-feelings}. As could be seen in the table, major tones induce happiness, in contrast of minor tones which feels sad, however, the role of musical intervals is not clear in this distinction.

\subsection{Music Theory}

All sound is explained by the phenomenon of vibration. The frequency of a vibration that produces sound is named as \textit{pitch}. Specific values of pitches are named as \textit{notes}. In western music, notes are represented by uppercase alphabet letters from A to G. 

The difference of pitch between two sound frequencies is measured in \textit{cents}. This logarithmic unit of measure is much more rich in resolution than it is needed when considering western music, as two notes at worst differ only by 100 cents, which is referred to as a \textit{semitone}. The interval between a note and the note double of its frequency is called an \textit{octave}. An octave is divided into twelve equal parts, which yields us all the available notes. Two notes an octave apart are represented by the same letter, but differentiated by their octave number (e.g., A4 and A5).

The intervals below an octave are the most significant in music, and in our study. A list of intervals, alongside its difference in semitones and its feeling is given in Table~\ref{tab:intervals}. An important feature of intervals is that each single one of them have a feeling associated with it, mainly differing as positive and negative. This observation is central to our hypothesis.

The concept of "key" in western music refers to the main note that feels like "resolution" or "stability" from the musical passage. This main note will usually define the relationship with all other notes, creating a hierarchy. A musical "scale" can be thought as the "blueprint" of the music, as alongside the key, the melody and harmony is arranged around a pre-determined key and scale. The amount of scales that could be used in a western music setting is quite numerous, though two of them are most commonly used: major and minor scale. The relevant aspect of these two scales is that the major scale feels happy and upbeat, while the minor scale feels sad and sentimental.

\begin{table}[b]
    \centering
    \caption{Intervals in an Octave}
    \label{tab:intervals}
    
    \begin{tabular}{| c | c |  c | c |}
        \hline
    
        \textbf{Note} & \textbf{$\Delta$Semitones} & \textbf{Interval Name} & \textbf{Feeling}  \\
        \hline \hline
        C & 0 & Unison & Neutral \\ \hline 
        C$\sharp$ & 1 & Minor Second & Unpleasant, Dark \\ \hline 
        D & 2 & Major Second & Neutral, Passing Tone \\ \hline
        D$\sharp$ & 3 & Minor Second & Tragedy, Sadness \\ \hline
        E & 4 & Major Second & Joy, Happiness \\ \hline
        F & 5 & Perfect Fourth & Buoyant, Neutral \\ \hline
        F$\sharp$ & 6 & Tritone & Danger, Devilishness \\ \hline
        G & 7 & Perfect Fifth & Stability \\ \hline
        G$\sharp$ & 8 & Minor Sixth & Anguish, Sadness \\ \hline
        A & 9 & Major Sixth & Winsomeness \\ \hline
        A$\sharp$ & 10 & Minor Seventh & Irresolution, Mournfulness \\ \hline
        B & 11 & Major Seventh & Aspiration \\ \hline
    \end{tabular}
\end{table}

\section{Methodology}

This section is dedicated to explain both the dataset, and the necessary details we have undertaken to test our hypothesis.

\subsection{Original Dataset}

Because the emotional labeling of musical passages are a highly subjective process~\cite{Li:2011:MDM:2016688}, we opted to use an existing dataset. We have used the dataset put together by Eerola et. al. first used in his 2009 study~\cite{eerola2009prediction}.  The passages all taken from film scores. The rationale for solely using film scores is given as their explicit goal of supporting the feeling at a particular moment in a film, thus it is easier to identify and associate with the emotional content of the passage.

There are a grand total of 360 passages, each approximately 15 seconds long, and given in \verb|mp3| format. To ensure the labeling is as accurate as possible, the original authors employed 12 music experts and 116 university students.

Each passage is concretely labeled as one of the 12 labels. 6 of these labels are comprised of basic emotions such as "happy", "sad", "tender", "anger", "fear", and "surprise". Rest 6 of the labels are associated with the 3D space of human emotions: \textit{valence}, \textit{tension}, and \textit{energy}. A separate label is used for high and low values for each dimension.

\subsection{Preprocessing}

For the sake of simplicity, we disregarded the 3D emotion models and focused on basic emotions. The initial survey of the remainder of the dataset indicated that passages labeled with anything other than \textit{happy} and \textit{sad} either did not possess a clear melodic structure that could be transcribed, or the main vector of emotion was not the melody but other factors such as tempo and loudness (or lack thereof). Filtering the dataset down to "happy" and "sad" samples yielded a total of 60 passages.

As we have mentioned before, the musical passages in the dataset are in \verb|mp3| format and the music contained other elements in addition to the melody, hence we needed to extract the melodies first. To do so, we have transcribed the melodies of the identified passages by ear and converted them to MIDI format. Since the only interest is in the impact of intervals, we have paid extra attention to the correctness of the pitch; timings and the velocity of each note has been transcribed on a best-effort basis. As part of this process, we also identified the key of the melody.

During the transcription, we identified that some of the passages, again, did not possess a clear melodic structure, and a few of them had multiple melodies. Omitting non-melodic passages and adding the extra melodies has yielded a dataset comprised of 49 records which is the final number of records we have conducted our experiments with.

To parse the MIDI files, we used a custom MIDI parser that read a MIDI file and returned only the stream of notes. Using this stream, we have derived several sets of features, which we will now explain in the next subsection.

\subsection{Feature Engineering}

Our main approach in deriving features is simply counting the intervals and averaging them with the length of the passage. There are two important details: how do we measure the distance the interval between two notes, and what is the reference point?

MIDI specifies a total of 128 notes, so technically a total of 128 intervals can be represented. However, in practice, intervals find themselves used up to thirteenths, with ones below octave used most commonly. Taking this into account, we limited the counted intervals up an octave, and considered intervals higher than an octave to be their counterpart below octave.

Since an interval is comprised of two notes, we need a reference point in the first step to calculate an interval. Four reference points have been considered:

\begin{enumerate}
    \item \textbf{Key Note}: This is the note that the passage resolves to. 
    \item \textbf{Preceding Note}: Given a note in a melody, calculate the interval with the next note.
    \item \textbf{First Note of Passage}: Consider the first note as the key. Western music often, but not always, starts with the key note.
    \item \textbf{Last Note of Passage}: Consider the last note as the key. Western music often, but not always, ends with the key note.
\end{enumerate}

The function that calculates the tonal difference for "key note" and "preceding note" features is slightly different. The difference mainly lies in that preceding note features counts descending intervals (first note's pitch is higher than the second one) separately whereas key note features do not. This is because the reference point in preceding note features are a specific note on an octave (e.g., C4), in contrast of key note features where the note does not refer to a specific note (e.g., key of C). In other words, the key note exists both under and over a given note, as long as there is consistency (i.e., always choosing the note under or over), the direction is not important. The exact procedures we used for tonal difference is listed in Algorithm \ref{alg:tonaldiff}. Important things to mention are: we used "the note under" for calculating absolute tonal diff, and considered the octave interval as a unison.
\begin{table*}[]
    \centering
    \caption{Feature Sets, Algorithms, and the Classification Accuracies}
    \label{tab:results}

    \begin{tabular}{|c|c|c|c|c|c|}
        \hline
         \backslashbox{\textbf{Features}}{\textbf{Algorithms}} & \textbf{Naive Bayes} & \textbf{kNN (k=4)} & \textbf{Weighted kNN (k=4, 1/distance)} & \textbf{SVM} & \textbf{Decision Tree (C4.5)} \\
         \hline \hline
         \textbf{Preceeding Note} & 50\% & 58\% & 56.25\% & 52\% & 37.5\% \\
         \hline
         \textbf{First Note as Key} & 66.66\% & 60.41\% & 64.58\% & 60.41\% & 62.5\% \\
         \hline
         \textbf{Last Note as Key} & 58.33\% & 47.91\% & 47.91\% & 58.33\% & 54.16\% \\
         \hline
         \textbf{Identified Key} & 77.08\% & 85.41\% & 83.33\% & 83.33\% & 79.16\% \\
         \hline
         
    \end{tabular}
\end{table*}

\begin{algorithm}
  \DontPrintSemicolon
  \SetKwFunction{FMain}{tonal\_diff}
  \SetKwProg{Fn}{function}{:}{end}
  
  \SetKwInOut{Input}{input}\SetKwInOut{Output}{output}
  
  \Input{Two notes, $a$ and $b$, represented as integers according to MIDI specification.}
  \Output{Difference in semitones.}
  
  \BlankLine
  
  \Fn{\FMain{$a$, $b$}}{
        \If{$a \geq b$} {
            \KwRet $(a - b) \bmod 12$
        } \If{$a < b$} {
            \KwRet $((a - b) \bmod 12) - 12$
        }
  }
  
  \BlankLine

  \SetKwFunction{FMain}{abs\_tonal\_diff}
  \SetKwProg{Pn}{function}{:}{end}
  \Pn{\FMain{$a$, $b$}}{
    \KwRet $(a - b) \bmod 12$
  }
  \label{alg:tonaldiff}
  \caption{Calculating tonal difference.}
\end{algorithm}

The final form of the data which we conducted our classification experiments are simply produced by iterating all of the notes to count the intervals with respect to the reference point. The key of the melody were identified by us and was available in the meta-data. All other features has been calculated by solely the notes in the passage.

\subsection{Classification}

The accuracy of the features and classification algorithms are given in Table~\ref{tab:results}. All the results have been obtained with 10-fold cross validation. Preceding-note and last-note-as-key features yielded almost random - or worse than random in case of a decision tree - classifiers. Classifiers using first-note-as-key features did slightly better than random. The highest accuracy was obtained by using the identified-key features used in conjunction with k-nearest-neighbor algorithm. We used WEKA \cite{hall09:_weka_data_minin_softw} to explore the data and conduct the experiments.

\section{Discussion}

The results came as a partial surprise. We initially expected features other than identified-key-features to be inferior to it indeed, but we expected them to perform much better than a random classifier. This indicates that the intervals between consecutive notes in a melody do not associate with the feeling.

The first-note-as-key features seem to perform slightly better than preceeding-note features, though we assert that the reason for this edge is that it is more likely for the first note to be the actual key of the passage. Last-note-as-key features did as bad as preceeding-note features, yielding random classifiers.

The identified key's performance can be explained by the differences in major and minor scale. Recall that music in major scale is often associated with positive emotions, and music in minor scale is often associated with negative emotions. Table~\ref{tab:interval_weights} shows the weights of individual intervals in a SVM classifier, trained/tested with 66\% percentage split and attained similar accuracy (approx. 81\%). Negative weights point towards "sad", and positive weights point towards "happy" labelings. From the table, it can be seen that the SVM model is in agreement with the overall feeling of major and minor scales, and thirds and sixths in particular are the most important intervals in emphasizing the feeling of the scale, and the passage.

It must be mentioned that even with identified-key features, the best classifier is rather far away from the perfect classifier (15\%). This gap can be explained if we consider even though major and minor scales are associated with happy and sad emotions respectively, composers often use notes outside the scale, or even change the scale of the music at an arbitrary point. 

By taking the failure of preceding-note features and the prevalence of identified-key features into account, we reason that in terms of emotional content, intervals themselves are only meaningful under correctly established musical context. This is in contrast with our initial expectations.

\begin{table}[]
    \caption{Weights of Individual Intervals in a SVM classifier}
    \label{tab:interval_weights}
    
    \centering
    \begin{tabular}{|c|c|c|}
        \hline
        \textbf{Interval} & \textbf{Appears in} & \textbf{Weight in SVM}  \\ \hline \hline
        Unison & Both & -0.36 \\ \hline
        Minor Second & None & -0.35 \\ \hline
        Major Second & Both & 0.03 \\ \hline
        Minor Third & Minor Scale & \textbf{-1.27} \\ \hline
        Major Third & Major Scale & \textbf{1.07} \\ \hline
        Perfect Fourth & Both & -0.18 \\ \hline
        Tritone & None & 0.24 \\ \hline
        Perfect Fifth & Both & 0.08 \\ \hline
        Minor Sixth & Minor Scale & \textbf{-1.26} \\ \hline
        Major Sixth & Major Scale & \textbf{2.07} \\ \hline
        Minor Seventh & Minor Scale & -0.21 \\ \hline
        Major Seventh & Major Scale & -0.11 \\ \hline
    \end{tabular}
\end{table}

\section{Conclusion}

In this study, we have examined the impact of intervals on the emotional effect specifically in western music. The utilized dataset has been constructed with the help of a large body of students and music experts in a 2009 study. We counted the intervals with respect to various reference points, averaged them by the length of the passage, and conducted classification on this derived dataset. Features derived by intervals among notes next to each other in a passage deemed ineffective, whereas counting intervals with respect to the key of the song yielded a relatively high accuracy classifier (approx. 85\%). We reason this indicates that intervals could only be associated with emotional dimension of music if they are evaluated in correctly established musical context.

\section{Related Work}

Most recent research is mainly focused on extracting emotional information directly from audio recordings, instead of notes. Much of this effort has been thorough MIREX (Music Information Retrieval EXchange) \cite{mirex}, which annually accepts and evaluates submissions on various musical data mining challenges, including mood classification, instrument recognition, automated drum transcription, and more. Most of the prevalent submissions utilize a form of convolutional neural network. Features extracted by the usage of Mel Frequency Cepstral Coefficients (MFCC) \cite{logan2000mel} is also popular among contestants.

On MIREX, the best accuracy for mood classification is around 69\%. One of the submissions that attained this figure \cite{lee2017cross} pre-trained their classifiers on a big tagged database for feature extraction, used two distinct deep neural network architecture outputting to an SVM to make a final decision. 

A few relatively dated publications regarding the use of notes for emotional classification is available. Wang et. al. \cite{wang2004recognition} used statistical features such as average and standard deviation of structs such as tempo, interval, pitch, loudness, and other MIDI and audio features to classify 6 emotions, grouped in three. The classification of final labels has been made in two steps, in the first step, an SVM decided whether the music was energetic or tranquil, and two separate SVM's each for the outcome for the first SVM was used to particularize the feeling, e.g., classifying between, joyous, robust, or restless in case the music is energetic. The accuracy for energetic-tranquil duo was around 95\%, while the accuracy for all 6 classes varied around 62\% and 85\%. 

Another study conducted by Lu et. al. \cite{lu2010boosting} used a plethora of features to classify between four emotions based on Thayer's arousal-valence emotion modeling. The features they used can be analyzed in three subjects: MIDI, audio, and lyrics. They used AdaBoost with SVM as the weak learning algorithm, and evaluated feature sets in all possible combinations (standalone, pairwise combination, and all of them). They found that using all the feature sets were superior to other combinations, and attained 72.4\% accuracy in that case.

The works mentioned here contains much merit, but none of them analyzed the musical intervals in context of emotional recognition as we have done in this study.

\bibliographystyle{IEEEtran}
\bibliography{references}

\end{document}